\input{aipcheck}

\documentclass[
    ,final            
  ]
  {aipproc}

\layoutstyle{6x9}

\begin{document}

\title{Neutrino-driven explosions twenty years after SN1987A}


\author{H.-Thomas Janka}{
  address={Max-Planck-Institute for Astrophysics, 
           Karl-Schwarzschild-Str.~1,
           D-85741 Garching, Germany}
}

\author{Andreas Marek}{
  address={Max-Planck-Institute for Astrophysics,  
           Karl-Schwarzschild-Str.~1, 
           D-85741 Garching, Germany}
}

\author{Francisco-Shu Kitaura}{
  address={Max-Planck-Institute for Astrophysics,  
           Karl-Schwarzschild-Str.~1, 
           D-85741 Garching, Germany}
}

\begin{abstract}
 The neutrino-heating mechanism remains a viable possibility for
 the cause of the explosion in a wide mass range of supernova
 progenitors. This is demonstrated by
 recent two-dimensional hydrodynamic simulations with detailed,
 energy-dependent neutrino transport. Neutrino-driven explosions
 were not only found for stars in the 8--10$\,M_\odot$
 range with ONeMg cores and in case of the iron core collapse
 of an 11$\,M_\odot$ progenitor, but also for a
 ``typical'' 15$\,M_\odot$ progenitor model. For such more massive
 stars, however, the explosion occurs significantly later than
 so far thought, and is crucially supported by large-amplitude
 bipolar oscillations due to the nonradial standing accretion shock
 instability (SASI), whose low (dipole and quadrupole) modes 
 can develop large growth rates in conditions where convective 
 instability is damped or even suppressed. The dominance of 
 low-mode deformation at the time of shock revival has been
 recognized as a possible explanation of large pulsar kicks and 
 of large-scale mixing phenomena observed in supernovae like SN~1987A.
\end{abstract}

\maketitle


\section{Introduction}

Besides the many other surprises Supernova~1987A brought for
astronomers, it had a major impact on the theory of stellar
core-collapse and explosion 
mainly by two discoveries. On the one hand the historical
detection of two dozen neutrino events in three underground
laboratories has confirmed the concept of gravitational
instability and neutron star formation, in which the production
of electron capture neutrinos and the emission of neutrinos and 
antineutrinos of all flavors by thermal processes had been 
predicted for a long time~\cite{Arnett.etal:1989}.
 
On the other hand, the lightcurve and spectra of SN~1987A brought
unambiguous evidence that nucleosynthesis products were distributed 
strongly anisotropically and that large-scale mixing took place during 
the explosion, for which reason X-rays and $\gamma$-rays from the 
decay of radioactive cobalt were measured much earlier than 
expected. Heavy elements were observed to expand 
with velocities significantly larger than expected from
spherically symmetric explosion models. This was interpreted as a
clear sign that the onion-shell structure of the progenitor star 
was destroyed during the explosion~\cite{Arnett.etal:1989}. 
Meanwhile, twenty
years later, the remnant of SN~1987A at the center of the ring
system reveals a clear prolate deformation and suggests a global
asymmetry of the mass ejection.

Multi-dimensional supernova models showed that sufficiently
strong radial mixing of radioactive nuclei requires that hydrodynamic
instabilities have developed in layers near the stellar core
and already during the earliest stages of the explosion.
In fact, simulations of 
the onset of the explosion demonstrated that strong convective
overturn can occur in the Ledoux-unstable region of neutrino
energy deposition behind the stalled supernova 
shock~\cite{Herant.etal:1994,Burrows.etal:1995,Janka.Mueller:1996}.

Meanwhile it is clear that convection is not the only source of
asymmetry during the shock stagnation phase. The standing
accretion shock has been recognized to be generically unstable to
nonradial deformation, even in situations where convection is 
damped or suppressed. This so-called ``standing accretion shock
instability'' (SASI; \cite{Blondin.etal:2003}; for more literature, 
see~\cite{Scheck.etal:2007}) shows a preferential growth of
low shock-deformation modes (dipole, $l=1$, and quadrupole, $l=2$, 
modes in terms of an expansion in spherical harmonics). The presence
of such a low-mode instability has turned out to have important
implications for large-scale explosion asymmetries, pulsar
kicks, and --- as suggested by very recent simulations --- for 
the development of neutrino-driven explosions. Corresponding results
will be reported below and implications for SN~1987A will be 
discussed.

\section{Explosion models with energy-dependent neutrino transport}

\subsection{Numerical method}

The core-collapse and post-bounce calculations presented here
were performed in spherical symmetry with the
neutrino-hydrodynamics code \textsc{Vertex}
(for details, see \cite{Rampp.Janka:2002,Buras.etal:2006}).
The code module that integrates the nonrelativistic hydrodynamics
equations is a conservative, Eulerian implementation of a
Godunov-type scheme with higher-order spatial and temporal accuracy.
The self-gravity of the stellar gas is treated with an
approximation to general relativity as described in \cite{Marek.etal:2006}.
The code was tested against fully relativistic simulations
in~\cite{Liebendoerfer.etal:2005,Marek.etal:2006}.
The time-implicit transport routine solves the moment
equations for neutrino number, energy, and momentum. It employs a
variable Eddington closure factor that is obtained from iterating
to convergence a simplified Boltzmann equation coupled to the set
of its moment equations.
The interactions of neutrinos ($\nu$) and antineutrinos ($\bar\nu$)
of all flavors include
a state-of-the-art treatment of charged-current and neutral-current
interactions with electrons, nucleons, and nuclei
(making use of the improved electron
capture rates on a very large NSE-ensemble of nuclei as considered by
\cite{Langanke.Martinez-Pinedo.ea:2003}). The most
important neutrino-pair processes in SNe as well as reactions
between neutrinos of different flavors are taken into
account~\cite{Buras.etal:2006,Marek.etal:2005}).

\begin{figure}
  \includegraphics[width=.48\textwidth]{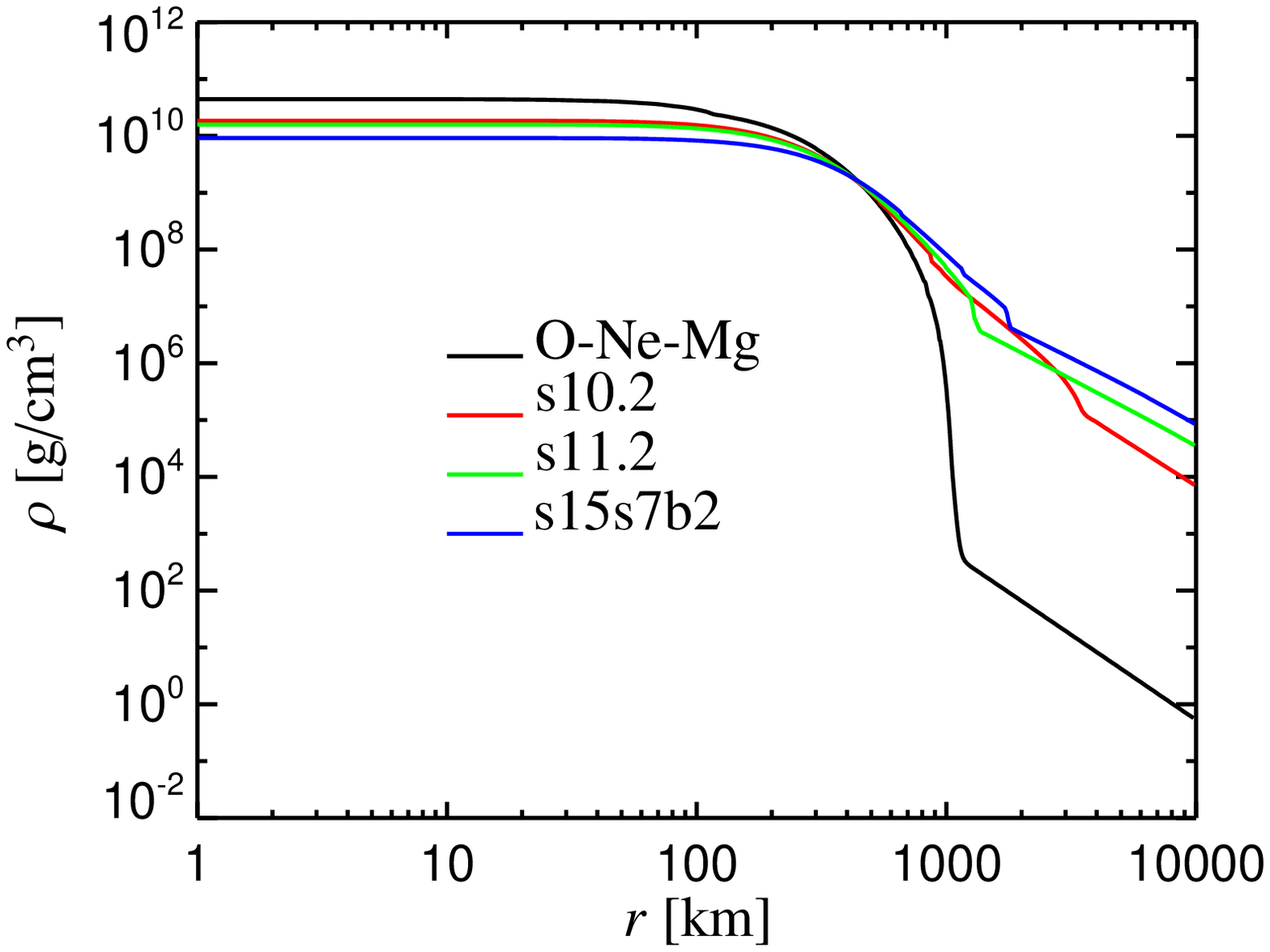}
  \includegraphics[width=.48\textwidth]{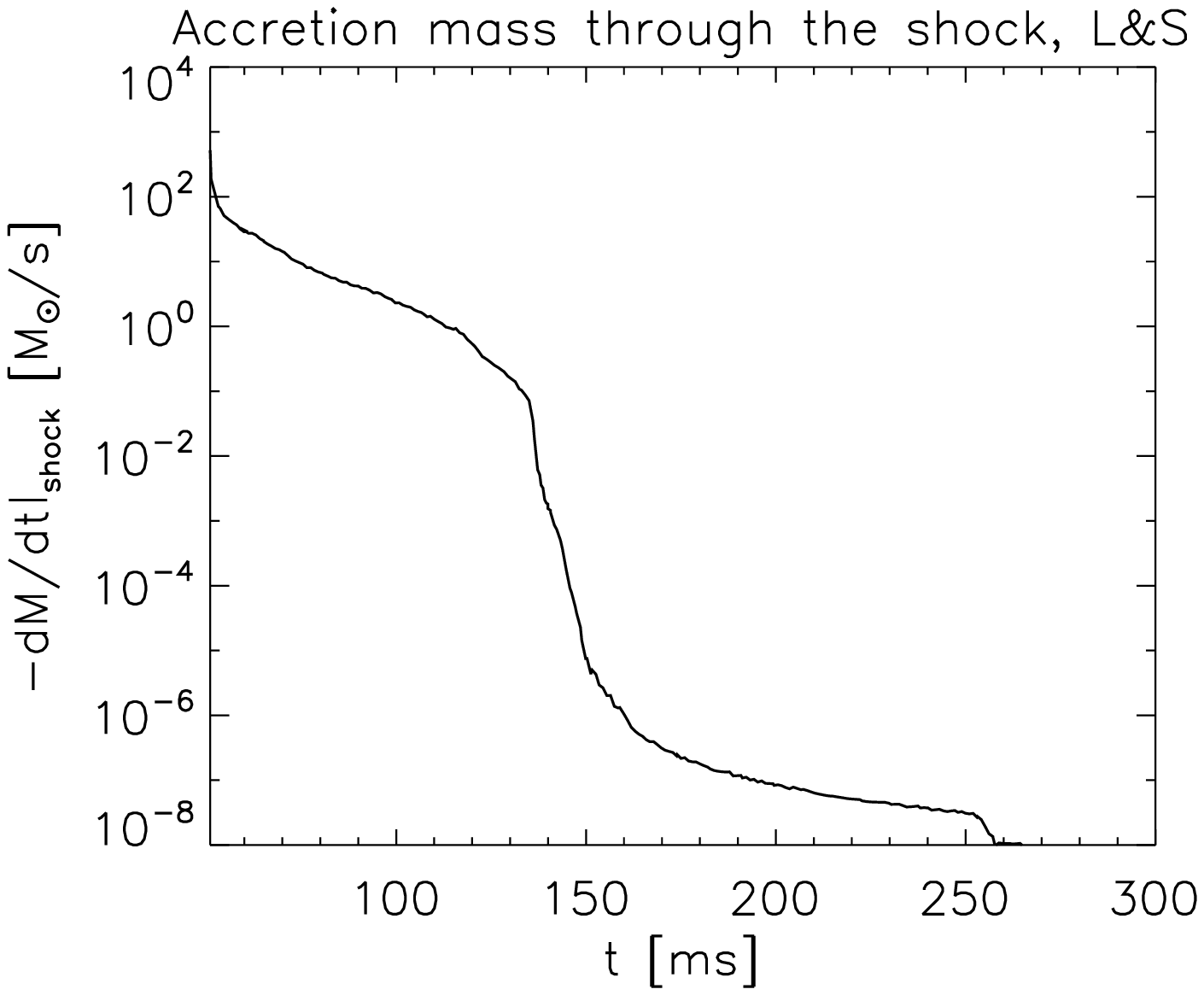}
  \caption{{\em Left:} Density profile of the ONeMg core and the
  surrounding He-shell of an 8.8$\,M_\odot$ star, which is 
  considered to be representative of the 8--10$\,M_\odot$
  range, compared to progenitor stars with 10.2, 11.2, and
  15$\,M_\odot$.
  Note that due to the lack of data from stellar evolution 
  models, the He-shell outside the oxygen-helium 
  transition at about 1000$\,$km was constructed from hydrostatic
  equilibrium, using a temperature profile as given by the 
  10.2$\,M_\odot$ progenitor (A.~Heger, private communication). 
  The actual density gradient is even
  steeper (K.~Nomoto, private communication). 
  {\em Right:} The mass accretion rate of the 
  collapsing ONeMg core at a function of time after bounce,
  measured just outside of the supernova shock }
\label{fig:ONeMgprofile}
\end{figure}

\begin{figure}
  \includegraphics[width=.50\textwidth]{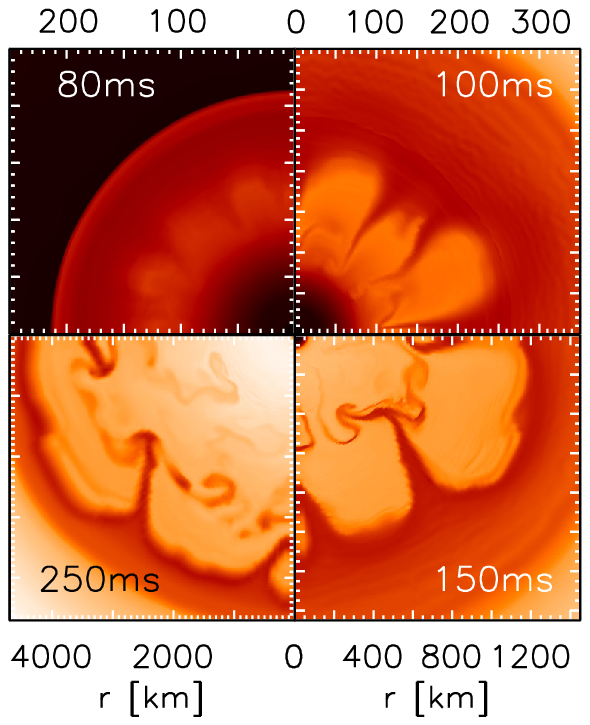}
  \includegraphics[width=.48\textwidth]{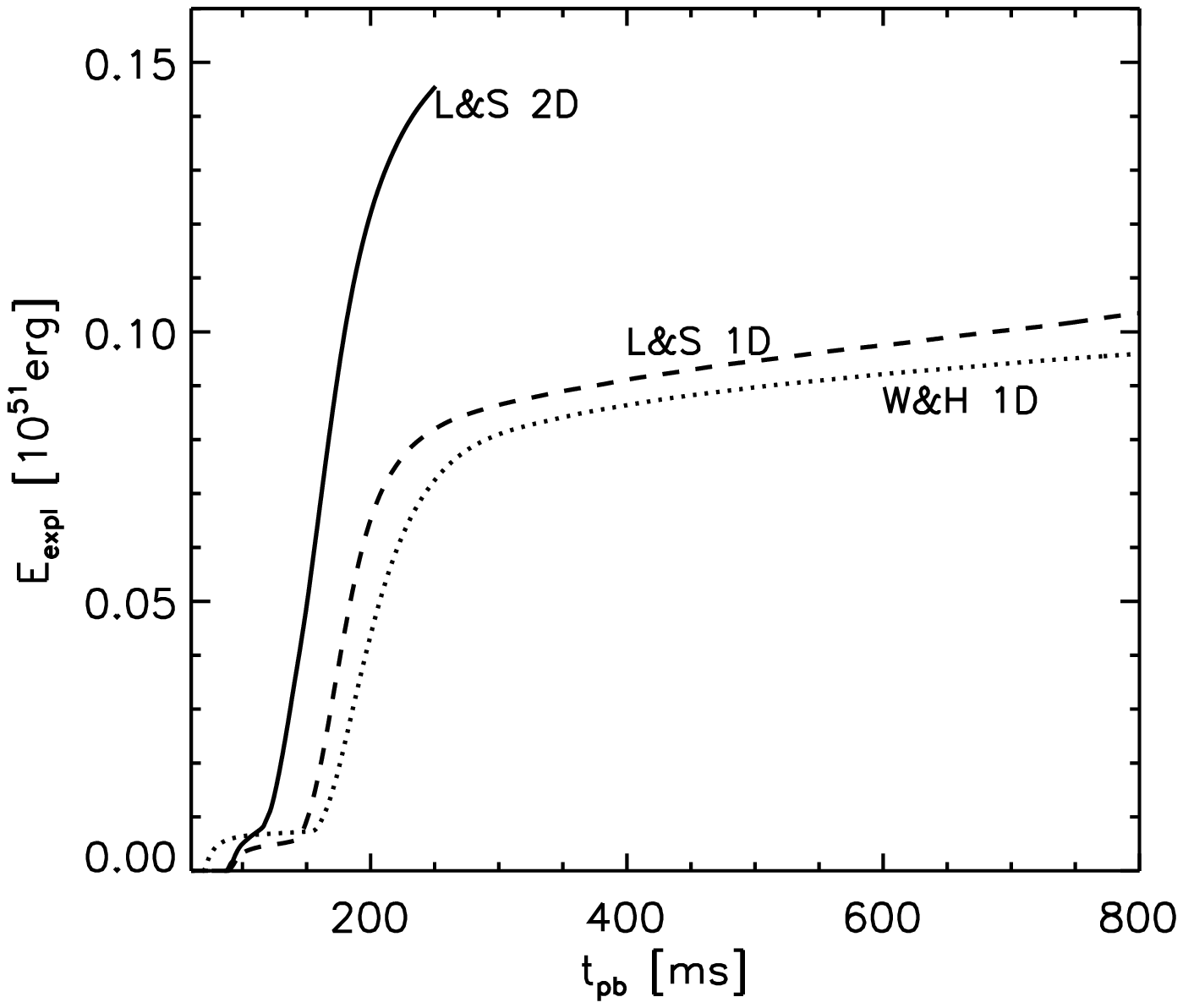}
  \caption{{\em Left:} Four snapshots of the explosion of an
   8--10$\,M_\odot$ star in a two-dimensional (2D) simulation, which
   was performed in a $\pm 45^\circ$ wedge around the 
   equatorial plane, using periodic boundary conditions. Time
   is normalized to bounce. The color coding represents the entropy
   per nucleon with black corresponding to values of 
   ${\,\hbox{\hbox{$ < $}\kern -0.8em \lower
   1.0ex\hbox{$\sim$}}\,}$7$\,k_\mathrm{B}$, red to 10--15$\,k_\mathrm{B}$,
   orange to 15--20$\,k_\mathrm{B}$, and white to about 
   25$\,k_\mathrm{B}$. The supernova shock is visible as
   sharp red/black discontinuity at about 210$\,$km 
   in the upper left panel, while
   it is already far outside the displayed region at all other
   times (the corresponding shock radii are roughly 900$\,$km,
   5600$\,$km, and 15000$\,$km).
   {\em Right:} Explosion energy as a function of time for the
   2D simulation of the left figure compared to two runs in
   spherical symmetry (1D) with
   a soft (``L\&S'') and a stiff (``W\&H'') nuclear equation of state.
   The steep increase of the explosion energy in the 1D models after
   about 150$\,$ms is caused by the onset of the expansion of 
   neutrino-heated matter away from the gain radius. 
   Convective overturn leads to more efficient neutrino
   heating of a larger mass and to an earlier rise of the explosion
   energy in the 2D simulation
   }
\label{fig:ONeMgexpl}
\end{figure}

\subsection{Neutrino-driven explosions for progenitors below 10$\,M_\odot$}

Recently Kitaura et al.~\cite{Kitaura.etal:2006} reinvestigated 
the stellar collapse of a 
$\sim$1.3$\,M_\odot$ core of oxygen, neon, and magnesium, surrounded 
by a thin ($\sim$0.08\,$M_\odot$) carbon layer and a very dilute helium shell.
The progenitor had 8.8$\,M_\odot$ on the main sequence mass and an 
initial He-core with 2.2$\,M_\odot$~\cite{Nomoto:1984}. 
It can be considered as
representative of the lowest-mass progenitors of core-collapse supernovae
in the 8--10$\,M_\odot$ range. 

Kitaura et al.\ obained an explosion that set in about 100$\,$ms after
core bounce and whose energy was provided by a neutrino-driven wind.
The spherically symmetric (1D) simulations confirm qualitatively older 
calculations by Mayle and Wilson~\cite{Mayle.Wilson:1988}, 
although the recent explosion
models are significantly less powerful and important differences
with respect to the nucleosynthesis conditions in the ejecta are seen.

Because of the presence of O, Ne, Mg, and C, nuclear burning still
proceeds in the outer regions of the stellar core while efficient 
electron capture (mostly on $^{20}$Ne, $^{24}$Na, and $^{24}$Mg) 
reduces the electron degeneracy pressure and drives the core to 
gravitational instability. 
It is, however, not the presence of the energy release by burning in
some shells that makes the explosion of stars with such cores
much easier than that of more massive progenitors with iron
in the center (the compressed matter in any case is heated to
nuclear statistical equilibrium, and the energy released by the
burning is efficiently removed by escaping neutrinos). The main
reason for the readiness of such low-mass stars to explode by the
neutrino-driven mechanism is the decreasing density in the C-layer 
and the extremely steep density gradient at the transition from the 
C-shell to the He-mantle (see the left plot in Fig.~\ref{fig:ONeMgprofile}). 
This leads to a continuous, fast drop of the mass accretion rate after
about 50$\,$ms of post-bounce evolution (Fig.~\ref{fig:ONeMgprofile},
right plot). As a consequence, the stalled prompt shock 
starts reexpanding and accelerates the very dilute matter in its
downstream region. At about 150$\,$ms after bounce material expands
outward from regions near the gain radius, where it was exposed to
intense neutrino heating. This phase is associated with a steep rise
of the explosion energy in Fig.~\ref{fig:ONeMgexpl} (right panel). 
Between 200 and 250$\,$ms after bounce a powerful neutrino-driven 
wind begins to shed off more gas from the surface of the nascent 
neutron star. From this time on the explosion energy in 
Fig.~\ref{fig:ONeMgexpl} shows a more gradual but continuous further
increase. 

Multi-dimensional effects are obviously not crucial for obtaining
neutrino-driven explosions of progenitors with the structure
of the considered $\sim\,$9$\,M_\odot$ model. Nevertheless, a 
simulation performed in two dimensions (2D; i.e., assuming axial 
symmetry) shows that convective overturn in the neutrino-heated layer 
between the gain radius (at 90$\,$km) and the shock becomes strong 
about 80$\,$ms after bounce and has fully developed 20$\,$ms later
(see Fig.~\ref{fig:ONeMgexpl}, upper panels of left plot). It 
carries cooler matter in narrow downdrafts from larger distances 
to locations closer to the gain radius, where the gas is
exposed to more efficient neutrino heating. Therefore a larger 
gas mass absorbs energy from neutrinos before it accelerates outward
in rising high-entropy plumes. This leads to a considerably higher
energy of the explosion than in the corresponding 1D simulations
(Fig.~\ref{fig:ONeMgexpl}, right plot), but has essentially no effect
on the propagation of the supernova shock during this phase, because
the shock is already far outside of the convective region. After 
about 150$\,$ms of post-bounce evolution the radial propagation of the 
neutrino-heated layers has become so fast that the mixing motions
freeze out and the corresponding fluid pattern with characteristic
Rayleigh-Taylor mushrooms expands self-similarly with high velocity
(Fig.~\ref{fig:ONeMgexpl}, lower panels in the left plot).

The 2D simulation also shows that convection inside the nascent
neutron star does not lead to any significant increase of the 
neutrino luminosities and thus of the neutrino heating behind the
shock. The enhanced explosion energy is merely a consequence of the
convective activity behind the supernova shock. This is 
clearly different from the 
simulations by Mayle \& Wilson~\cite{Mayle.Wilson:1988}, 
who obtained models with larger
explosion energy by assuming that the neutrino luminosities were
boosted by neutron-finger convection below the neutrinosphere.

The rapid outward acceleration also has the consequence that the
convective pattern never develops dominant power on the largest scales.
The expansion of the gain layer happens so quickly that the convective
plumes have no time to merge to structures with lateral wavelengths
of more than about 45$^\circ$. Since the shock radius grows continuously
with time, also the SASI has no possibility to grow (for more details,
see below). Such a situation disfavors the development of a large global 
asymmetry of the small amount of material that is accelerated during the 
early stages of the explosion. Therefore the pulsar kick velocities must
be expected to remain rather small 
(roughly ${\,\hbox{\hbox{$ < $}\kern -0.8em \lower
1.0ex\hbox{$\sim$}}\,}$100$\,$km/s) in case of the O-Ne-Mg core collapse
events.

\begin{figure}
  \includegraphics[width=.48\textwidth]{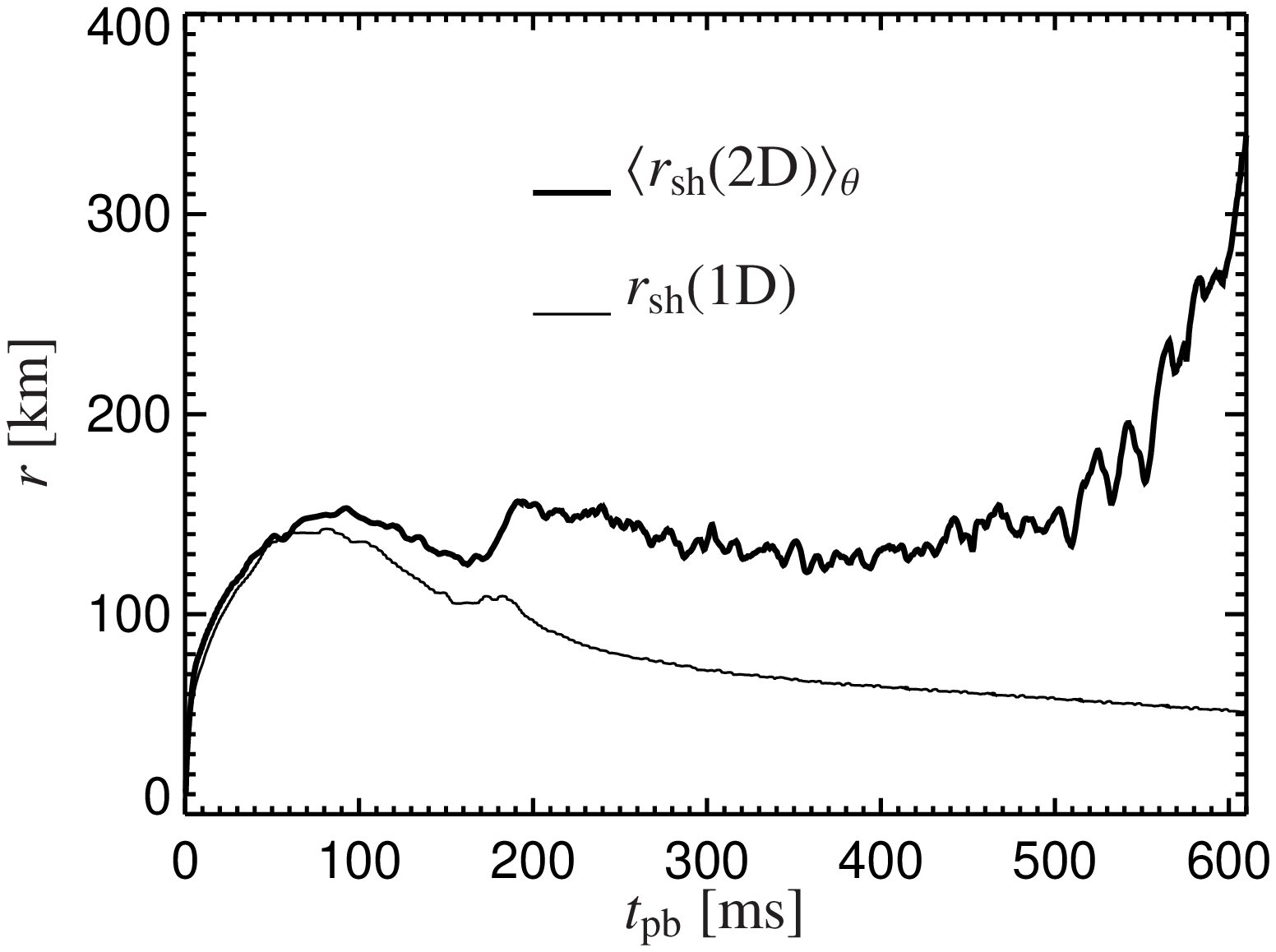}
  \includegraphics[width=.48\textwidth]{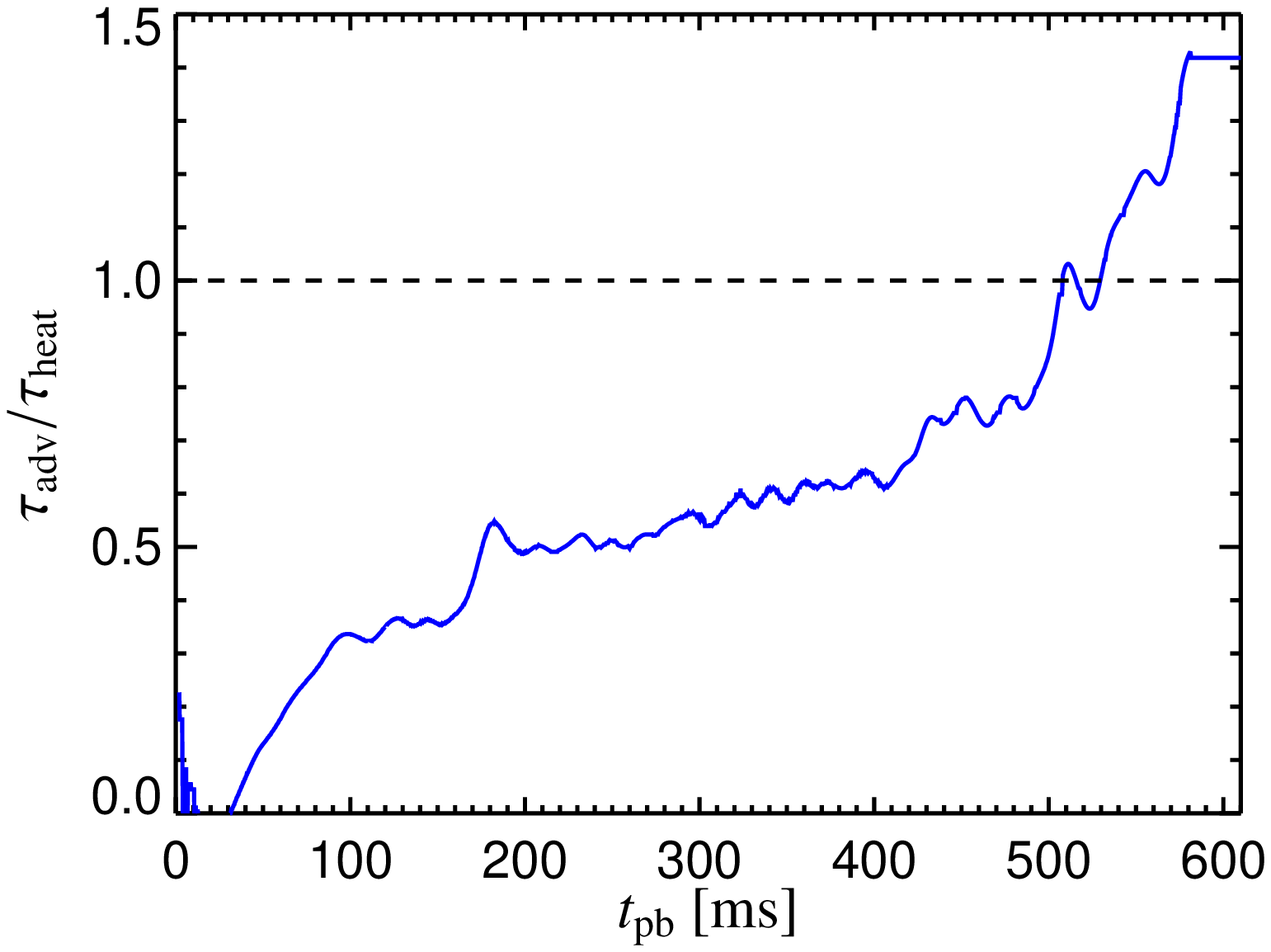}
  \caption{{\em Left:} Angular average of the shock radius (volume
   weighted) for the 2D
   simulation of the 15$\,M_\odot$ explosion compared to the
   shock position of a corresponding spherically symmetric simulation.
   {\em Right:} Evolution of the ratio of advection timescale of
   accreted matter through the gain layer to the neutrino-heating
   timescale
   for the exploding 15$\,M_\odot$ model. There is a continuous
   increase until the critical value of unity is exceeded after about
   500$\,$ms of post-bounce evolution. At $t>580\,$ms the beginning
   strong overall expansion of the postshock layer prevents a
   reasonable determination of the advection timescale }
\label{fig:15Msunexpl-rst}
\end{figure}

\begin{figure}
  \includegraphics[width=.99\textwidth]{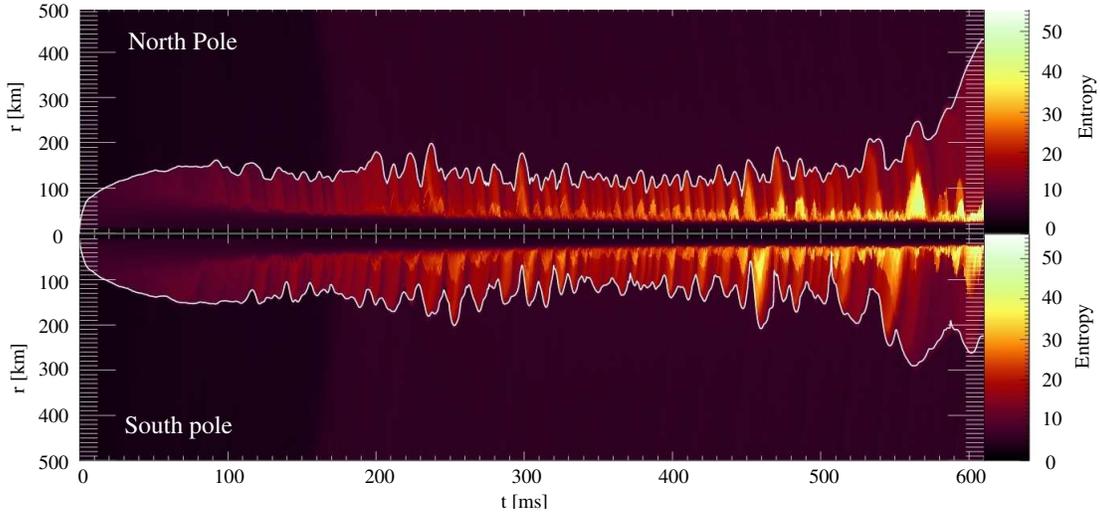}
  \caption{Radial positions of the shock near the north and south pole
   as functions of post-bounce time (white lines) 
   in the 2D simulation of the explosion of a
   15$\,M_\odot$ star. The color
   coding represents the entropy per nucleon of the stellar gas.
   The quasi-periodic shock expansion and contraction due to the
   SASI can be clearly seen }
\label{fig:15Msunexpl-rse}
\end{figure}

\begin{figure}
\tabcolsep=0.5mm
\begin{tabular}{lr}
  \includegraphics[width=.48\textwidth]{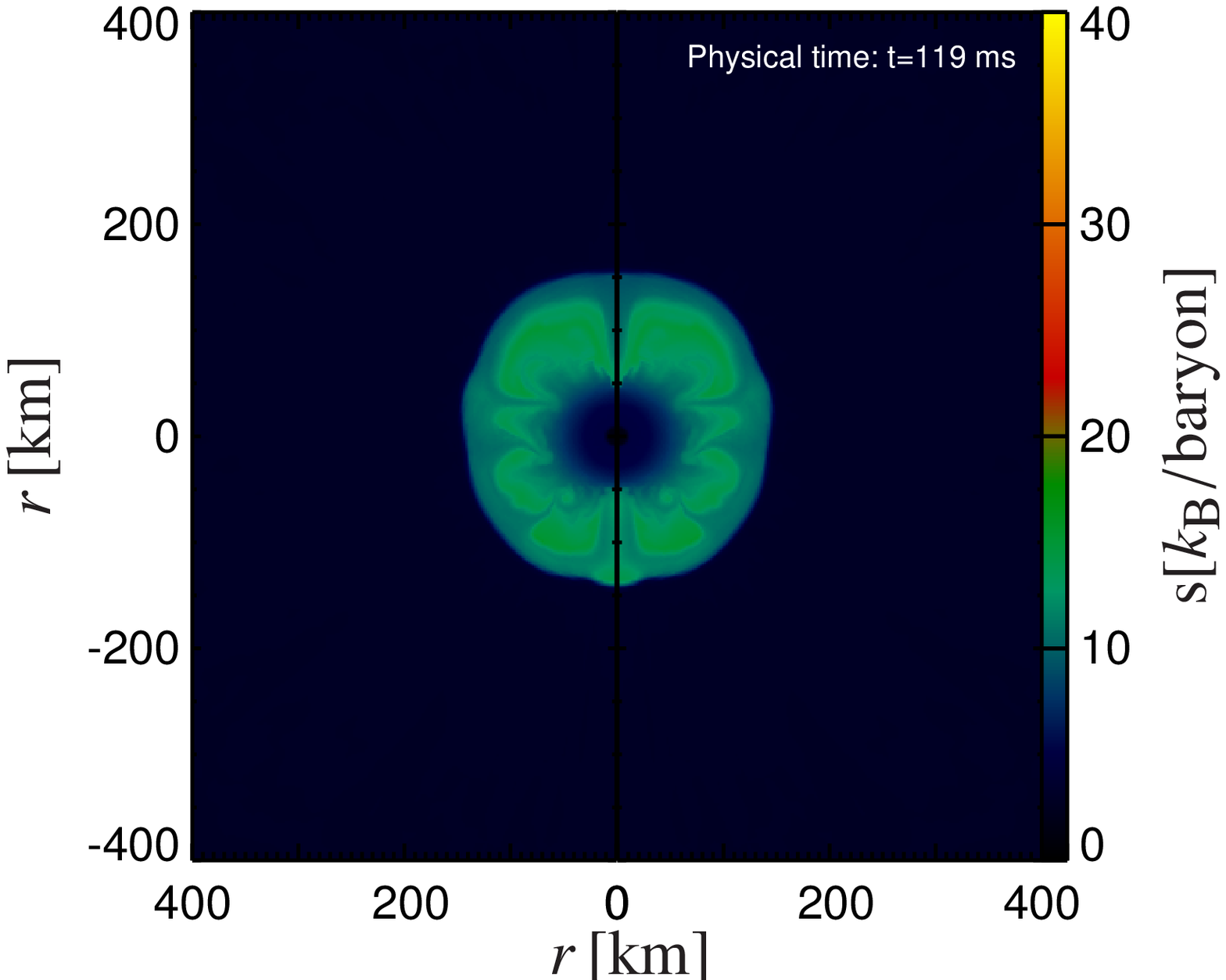} &
  \includegraphics[width=.48\textwidth]{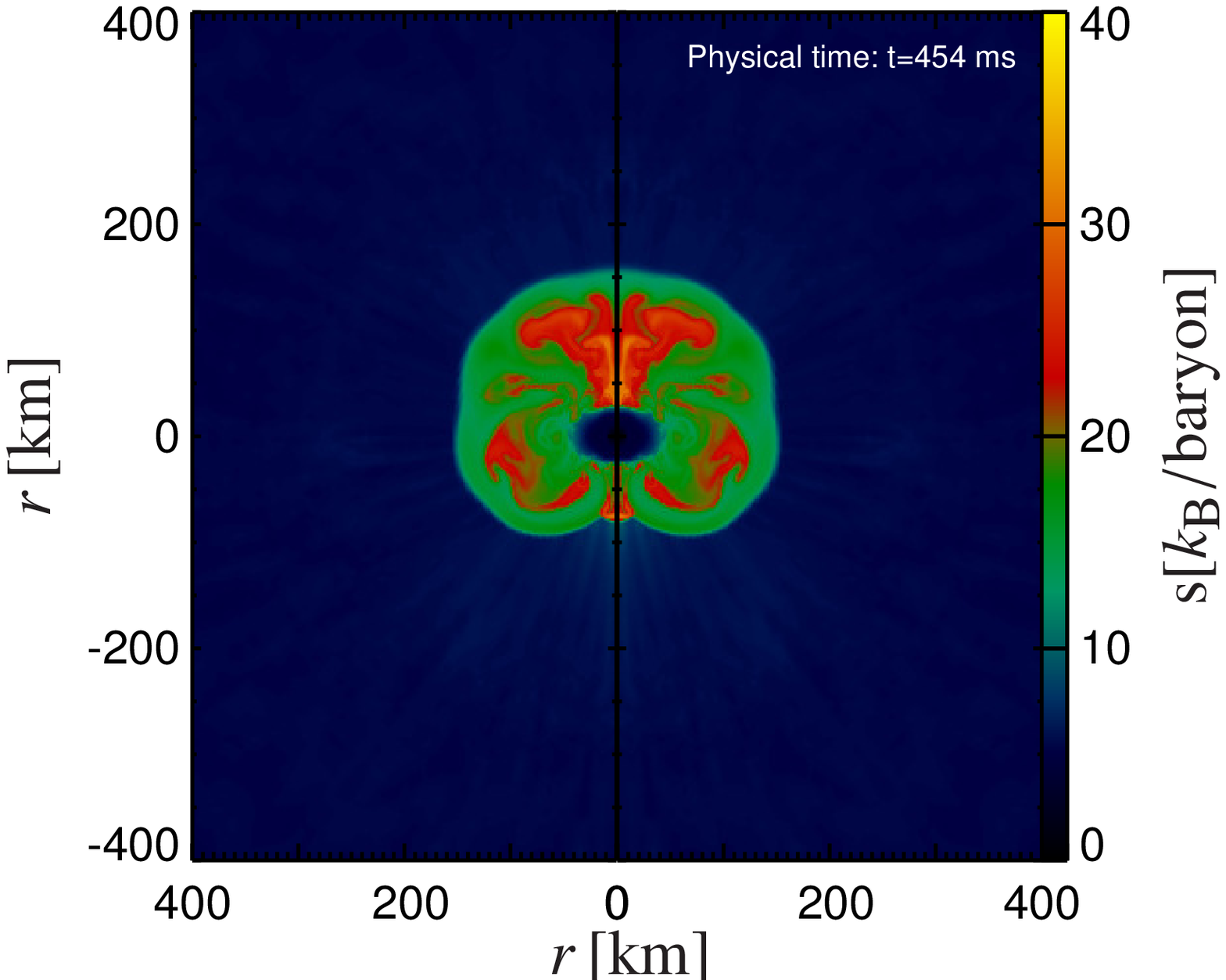}\\
  \includegraphics[width=.48\textwidth]{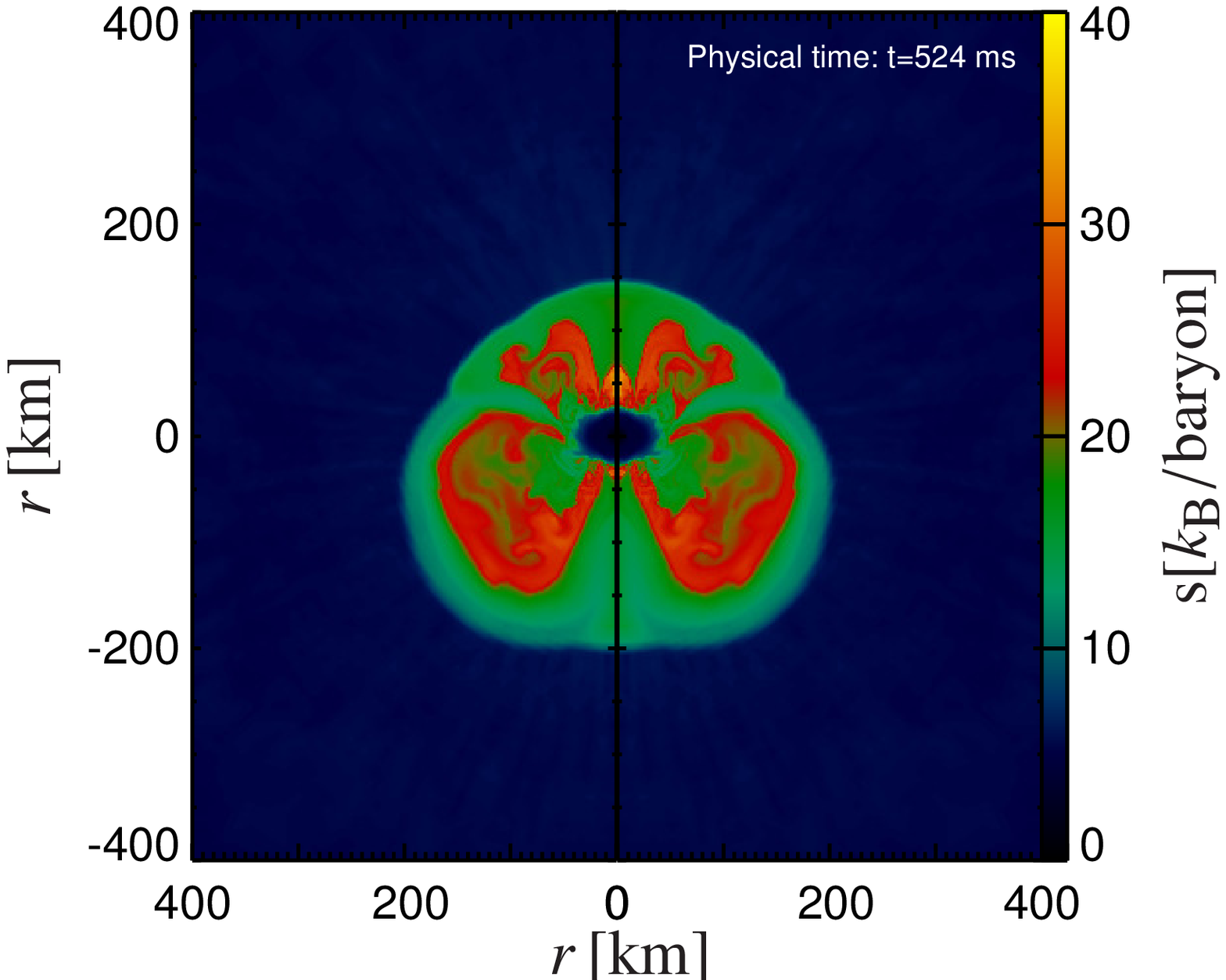} &
  \includegraphics[width=.48\textwidth]{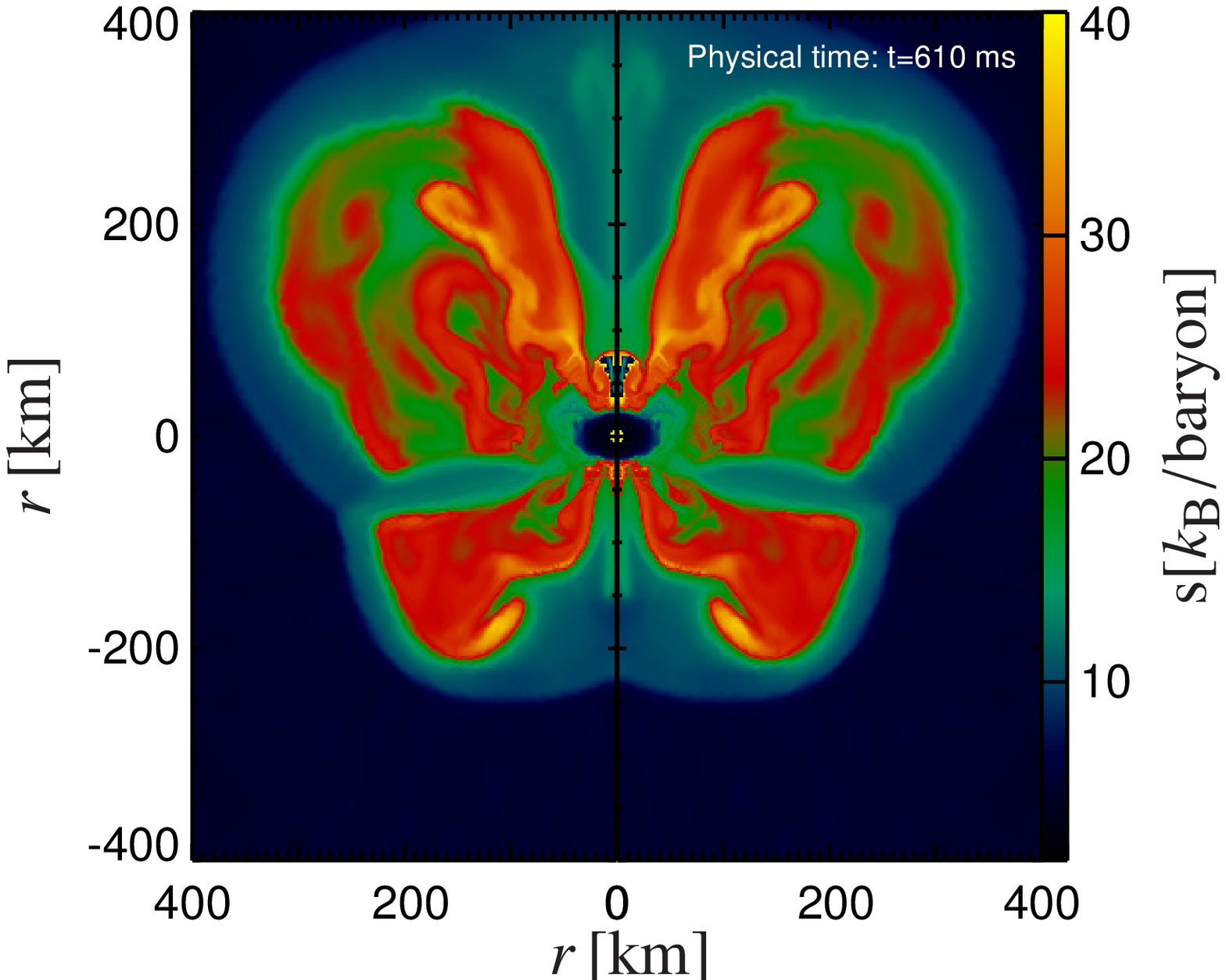}
\end{tabular}
  \caption{Four snapshots from the post-bounce
  evolution of the exploding 15$\,M_\odot$ star in a 2D simulation. 
  The upper left plot
  shows the entropy distribution at $t= 119\,$ms after bounce,
  about 40$\,$ms after the postshock convection has reached the
  nonlinear regime. The upper right and lower left plots ($t = 454\,$ms
  and $t = 524\,$ms after bounce) demonstrate the presence of a
  very strong bipolar oscillation due to the SASI, and the lower
  right plot ($t = 610\,$ms p.b.) displays the acceleration phase
  of the strongly aspherical explosion with a large $l=1$
  component. Note that the contracting nascent neutron star 
  exhibits a growing prolate deformation because of the rotation
  assumed in this simulation }
\label{fig:15Msunexpl-snaps}
\end{figure}

\subsection{SASI-supported neutrino-driven explosions of stars 
above 10$\,M_\odot$}

The core structure of stars more massive than about 10$\,M_\odot$
is considerably different from that of lower mass supernova 
progenitors (see Fig.~\ref{fig:ONeMgprofile}). Spherically symmetric
calculations, carried out over many hundreds of milliseconds after
core bounce, have therefore not found explosions happening. Instead,
the supernova shock stalls and mass is continuously accreting onto
the forming neutron star (see the 1D result in the left
plot of Fig.~\ref{fig:15Msunexpl-rst}).

Hydrodynamic instabilities in the supernova core, however,
can change the situation. In 2D simulations
Buras et al.~\cite{Buras.etal:2006b} obtained an explosion 
of an 11.2$\,M_\odot$ progenitor. Numerical tests with
different angular wedges and lateral boundary conditions of the 
polar grid showed that the crucial difference here was the 
growth of low ($l = 1,\,2$) SASI modes. The associated development
of large-amplitude bipolar oscillations pushed the shock to larger
radii and thus increased the timescale of accreted matter to fall from
the shock (at $R_\mathrm{s}$) to the gain radius $R_\mathrm{g}$. 
The corresponding advection timescale
\begin{equation}
\tau_\mathrm{adv}\, \equiv\, {R_\mathrm{s} - R_\mathrm{g} 
\over |\left\langle v_r \right\rangle|}
\end{equation}
can be considered as a measure of the duration gas
is exposed to neutrino heating in the gain layer. When the stalled
shock reaches a larger radius $R_\mathrm{s}$, the preshock velocity and 
average postshock velocity $\left\langle v_r \right\rangle$ are 
significantly smaller, which leads to a
considerably longer advection timescale (roughly $\tau_\mathrm{adv}
\propto R_\mathrm{s}^{3/2}$; Eq.~(15) in~\cite{Janka.etal:2001}). 
Our numerical
experiments showed that the presence of postshock convection alone 
(if the low SASI modes were suppressed by grid constraints)
was unable to provide enough support for a neutrino-driven 
explosion. When SASI oscillations helped increasing the shock radius,
however, the crucial ratio of advection timescale to neutrino heating
timescale grows and finally exceeds the critical value of unity. The
neutrino heating timescale,
\begin{equation}
\tau_\mathrm{heat}\,\equiv\, 
{E_\mathrm{bind}[R_\mathrm{gain},\,R_\mathrm{shock}] \over
                      Q_\mathrm{heat}}
\end{equation}
measures the time
it takes neutrinos to deposit (with an integrated rate
$Q_\mathrm{heat}$) and energy equal to the binding energy
$E_\mathrm{bind}[R_\mathrm{gain},\,R_\mathrm{shock}]$ of the matter 
in the gain layer.

Very recent simulations show that such a positive feedback 
between low-mode SASI oscillations and neutrino heating also occurs
in a 15$\,M_\odot$ progenitor (model s15s7b2 of~\cite{Woosley.Weaver:1995}). 
Also there it finally leads to a neutrino-driven explosion,
however at much a later time after core bounce 
(Fig.~\ref{fig:15Msunexpl-rst}). The particular model considered
here includes a modest amount of rotation (the pre-collapse
iron core had a rotation period of about 12~seconds as in 
Sect.~3.4 of~\cite{Buras.etal:2006b}), which explains a growing oblateness
of the nascent neutron star (see Fig.~\ref{fig:15Msunexpl-snaps}). 
Comparison with non-rotating models, however, reveals that 
angular momentum dependent effects may cause some quantitative
differences (and may to some extent foster the evolution towards 
an explosion) but do not seem to be the essential ingredient that
determines the overall behavior of the collapsing stellar core
in the long run\footnote{Because
of the considerable CPU-time requirements of 2D simulations 
with our sophisticated, energy-dependent neutrino transport, 
we could not yet carry the comparative runs of non-rotating
models to the very late post-bounce
time reached in the case presented here.}.

Figure~\ref{fig:15Msunexpl-rst} (left) reveals a growth of the 
average shock radius, which starts at about 350$\,$ms after
bounce and is accompanied by a continuous rise of the timescale
ratio $\tau_\mathrm{adv}/\tau_\mathrm{heat}$ 
(Fig.~\ref{fig:15Msunexpl-rst}, right). This rise
is caused by an increase of the average advection timescale
$\tau_\mathrm{adv}$, while $\tau_\mathrm{heat}$
remains nearly constant. The kinetic energy (also for the lateral
component of the velocity) in the gain layer
triples during this period of the evolution (while the rotational
energy changes only by a modest amount), suggesting that nonradial
fluid motions become more and more violent during this phase. 
Indeed, the bipolar SASI oscillations, which are visible from alternating
shock expansion and contraction phases in the northern and southern 
hemispheres with a period of 10--15$\,$ms, exhibit a growing
amplitude for $t_\mathrm{pb} > 350\,$ms (Fig.~\ref{fig:15Msunexpl-rse}).
With a larger average shock radius also more mass is accumulated
in the gain layer. At $t {\,\hbox{\hbox{$ > $}\kern -0.8em \lower
1.0ex\hbox{$\sim$}}\,}530\,$ms the critical timescale ratio 
exceeds unity and a runaway situation is reached. The accelerating 
overall expansion indicates the onset of a strongly aspherical,
neutrino-powered explosion (Fig.~\ref{fig:15Msunexpl-snaps}).

\begin{figure}
  \includegraphics[width=.32\textwidth]{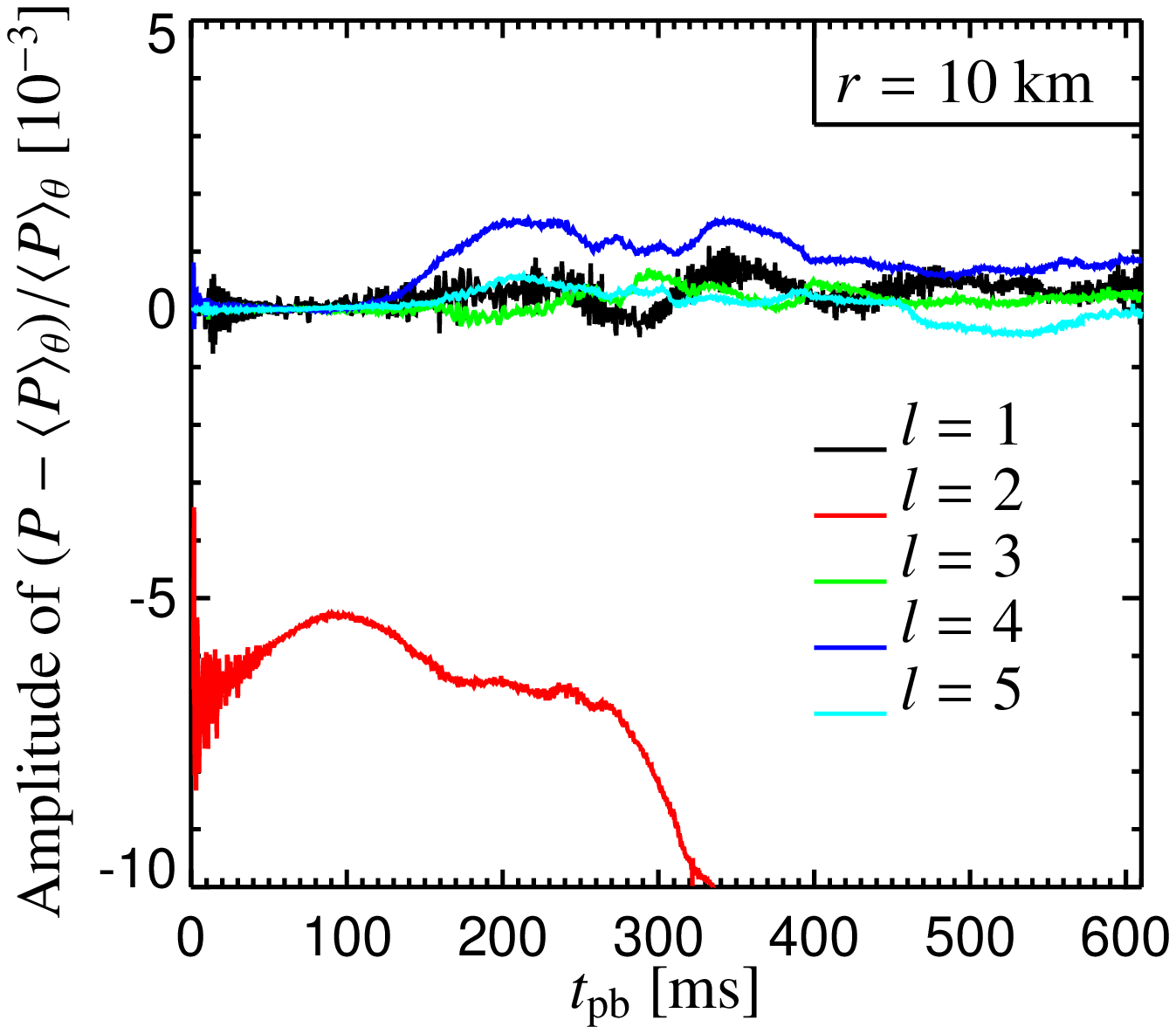}
  \includegraphics[width=.32\textwidth]{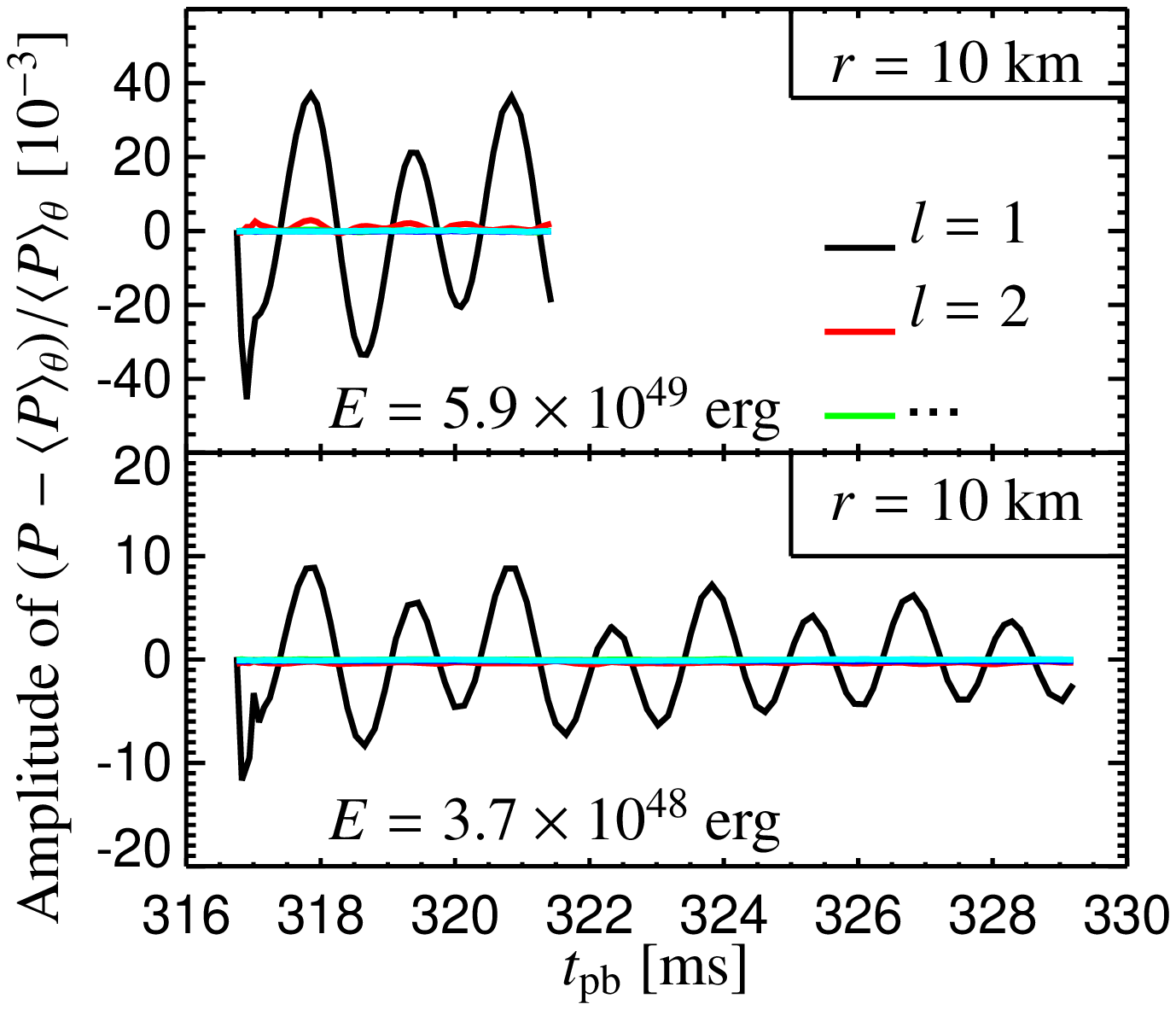}
  \includegraphics[width=.32\textwidth]{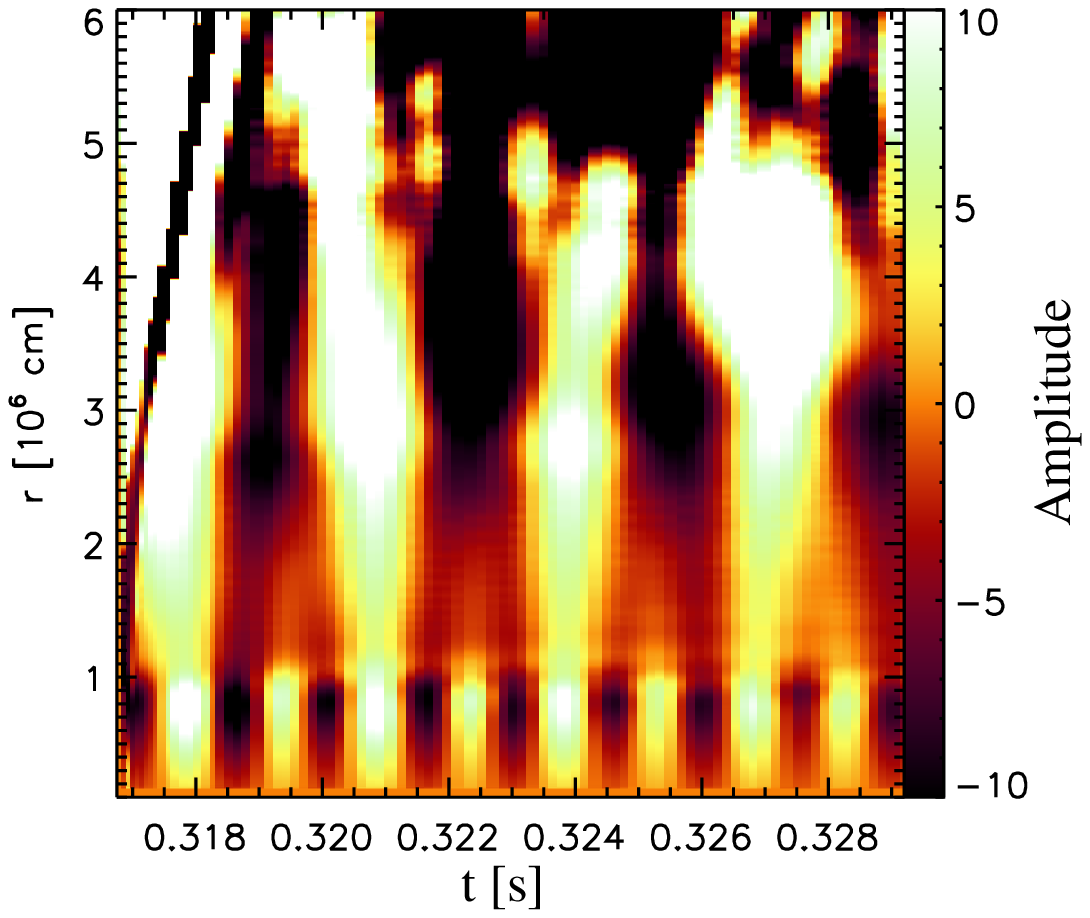}
  \caption{{\em Left:} G-mode oscillations of the
  nascent neutron star in the exploding 15$\,M_\odot$ simulation
  during 610$\,$ms of post-bounce evolution. The plot shows the
  amplitudes of the $l=1$ to $l=5$ modes of the pressure
  fluctuations at a radius of $r = 10\,$km expanded in spherical 
  harmonics. Note that the quadrupole mode ($l=2$) has a large
  and growing amplitude because of the oblateness of the rotating
  neutron star.
  {\em Middle:} Test simulations with artificially
  instigated dipole ($l=1$) oscillation of the neutron star. Two
  different amplitudes of the initially imposed velocity field
  were used, $5\times 10^7\,$cm/s and $2\times 10^8\,$cm/s,
  corresponding to a factor of 16 different kinetic energies (as
  indicated in the plot). The clear presence of many cycles of
  the dipole oscillation
  demonstrates the ability of our numerical code to follow such
  gravity waves, if they are excited.
  {\em Right:} The amplitude of the $l=1$ mode in the lower panel of
  the middle plot as function of time and radius. Interior
  of about 10$\,$km the core oscillates with twice the frequency as the
  mantle outside of $r\approx 25\,$km. In the intermediate, convective
  layer the gravity waves are damped }
\label{fig:15Msun-NSosc}
\end{figure}

\subsection{Some comments on core g-modes and the acoustic 
mechanism}

In view of the recent numerical finding of 
acoustically-driven explosions,
which are initiated by the acoustic power generated by 
large-amplitude core g-mode oscillations of the accreting 
neutron star~\cite{Burrows.etal:2006,Burrows.etal:2007}, 
we have evaluated our long-time 
15$\,M_\odot$ simulation for the gravity-wave activity of
the forming compact remnant. Figure~\ref{fig:15Msun-NSosc}
(left) displays the g-mode amplitudes of the first terms
($l=1,...,5$) of a spherical harmonics expansion of the 
pressure fluctuations at a radius of 10$\,$km inside the 
neutron star. The analysis follows the description in 
Ref.~\cite{Burrows.etal:2006},
see Fig.~7 there. The amplitudes of core g-modes in our model
are roughly two orders of magnitude smaller than those seen 
in the run-up to an explosion in that figure. The acoustic energy
flux radiated by the oscillating neutron star in our model
is therefore completely negligible compared to neutrino heating
behind the shock, which typically deposits energy at a net rate of
3--4$\times 10^{51}\,$erg/s at $t > 200\,$ms after bounce.
The acoustic mechanism does not play a role for the
evolution of our model and, according to the simulations
in\cite{Burrows.etal:2007}, it might become relevant only 
much later than our model explodes by neutrino-energy deposition.

But is our code able to follow core g-mode oscillations,
in particular of $l=1$ type, because in this case the gas in
the stellar center participates in the motion?
The answer is ``yes'' (in contrast to
statements that can be found in the literature\footnote{It is
true that in our simulations a few radial zones in the
central $\sim$1.5$\,$km of the star are treated in spherical
symmetry to get around the most severe CFL constraint for the
hydrodynamic timestep. This small central region within a 
protoneutron star of radius 15--50$\,$km, however, resembles
a pinhead in the middle of a cup filled with sloshing tea.}, 
see~\cite{Burrows.etal:2006,Burrows.etal:2007}).
The middle and right panels of Figure~\ref{fig:15Msun-NSosc}
show results of test simulations in which at some moment of
the post-bounce evolution we artificially instigated a large
dipole g-mode by imposing an $l=1,\,n=1$ (i.e., we assumed one
radial node) perturbation of the $z$-component of the velocity
field with varied amplitude and conserved linear momentum. The plots
demonstrate that essentially a pure $l=1$ oscillation develops
(after some initial relaxation, because our chosen perturbation
did not correspond to an eigenfunction), which the code is
able to follow through many cycles. We are therefore confident
that we should see large core g-mode oscillations, if the 
anisotropic accretion flow around the neutron star were causing 
their excitation.

\section{Conclusions}

The results of 2D supernova simulations presented in this paper 
demonstrate the ability of neutrino heating to initiate
delayed explosions for progenitors in a wider range of masses.
The explosion occurs significantly later than observed in
older calculations with approximative neutrino transport.
We identified large-amplitude SASI modes to play a crucial,
supportive role for the development of the explosion because
they enforce shock expansion and thus reduce the average infall
velocity in the postshock region, which enables the accreted 
matter to stay in the neutrino-heating layer for a significantly
longer time.

Our simulations, however,
were stopped too early (for CPU time reasons) to allow for a final
determination of the explosion energy. Accretion of matter by the 
shock is still going on, in particular in the 11 and 15$\,M_\odot$ 
stars, and gas is channelled towards the gain radius, where neutrino
heating is strongest. A large
fraction of this infalling material will start reexpanding, and
energy this gas has absorbed from neutrinos and is released by 
nucleon recombination to alpha particles and iron-group nuclei
will contribute to the explosion energy.
In order to obtain reliable numbers for the explosion
properties, the simulations will have to follow this accretion 
phase, which might last even for hundreds of milliseconds.
Ultimately, however, 3D simulations will be needed. 
The explosion, its onset
and strength, may depend on the additional degrees of freedom that
are accessible to the fluid flow in three dimensions. 
Convective downdrafts and buoyant plumes,
vorticity, and spiral modes are different in 3D or even do not exist
when the flow is constrained to axisymmetry with all structures 
being tori around the polar grid axis.

The kind of asphericities seen in case of our 11.2 and 15$\,M_\odot$
explosion models, with a large contribution from an $l=1$ component,
were shown to lead to such a big anisotropy of the supernova mass
ejection that the neutron star receives a recoil sufficiently strong
to explain the high velocities observed for many young pulsars,
even those in excess of 
1000$\,$km/s~\cite{Scheck.etal:2004,Scheck.etal:2006}. 
Moreover, the initial deformation of the supernova shock and the asymmetric
ejecta distribution are the seed of subsequent hydrodynamic instabilities
at the composition interfaces of the disrupted star after the passage 
of the supernova shock. These instabilities prevent the strong
deceleration of the heavy elements and lead to a highly anisotropic 
distribution not only of Fe-group nuclei but also of silicon and 
oxygen. Large-scale mixing takes place, in course of which hydrogen
and helium are carried deep into the star and pockets and clumps
of heavy elements remain expanding with high velocities as 
observed in SN~1987A~\cite{Kifonidis.etal:2006}.

Even 20 years after the spectacular stellar death it is not clear what 
caused the
explosion of SN~1987A. The ring system was interpreted as a 
sign for rapid rotation being present in the $\sim$18$\,M_\odot$
progenitor star. In particular the existence of a common
axis of the ring system and of the elongated ejecta is a strong
indication that rotation has played a role in the dying star,
possibly as the consequence of a binary merger event some
ten thousand years before the stellar collapse (see P.~Podsiadlowski's
talk at this meeting). It is, however, not clear how such a merger 
has affected the angular momentum evolution of the stellar core. Only 
if the initial spin period of the core was small 
(${\,\hbox{\hbox{$ < $}\kern -0.8em \lower 1.0ex\hbox{$\sim$}}\,} 2\,$s
according to Ref.~\cite{Burrows.etal:2007b}), 
the free energy of rotation in the nascent 
neutron star was sufficiently large to power a supernova
explosion by magnetohydrodynamic effects. But if the collapsing core was 
rotating so rapidly, why then
is there no sign now of the energy input from a bright, Crab-like pulsar?
A delayed collapse of a transiently existing neutron
star to a black hole is disfavored as the solution of this puzzle,
because the compact remnant formed in a typical 
SN~1987A progenitor is not expected to be so heavy that it cannot
be stabilized by nuclear equation-of-states that are consistent with
measured neutron star masses. 
Moreover, the pronounced prolate deformation of the 
now visible supernova ejecta at the center of the ring system may not be
an unambiguous signature of very rapid core rotation but could result from a
bipolar SASI asymmetry. 
SN~1987A may not only have been a unique event, it may also have been
an uncommon one. We will probably never find out with final certainty. 
The next
galactic supernova, however, will give us a new chance to learn
more about the processes that trigger the explosion of a massive star: 
Tens of thousands of neutrino events will be captured by various
underground experiments, and highly sensitive instruments promise 
to register the gravitational-wave signal produced by a nonspherical
bounce and by hydrodynamic instabilities in the supernova core.

\begin{theacknowledgments}
  We are very grateful to R.~Buras, W.~Hillebrandt, K.~Kifonidis, 
  B.~M\"uller, E.~M\"uller, and M.~Rampp for their input to various
  aspects of the reported project, and A.~Heger, K.~Nomoto, and 
  S.~Woosley for data of their progenitor models.
  This work was supported by the Deutsche Forschungsgemeinschaft
  through the Transregional Collaborative Research Center SFB/TR~27 
  ``Neutrinos and Beyond'', the Collaborative Research Center
  SFB-375 ``Astro-Particle Physics'', and the Cluster of Excellence 
  ``Origin and Structure of the Universe'' 
  (\url{http://www.universe-cluster.de}). Supercomputer time grants 
  at the John von Neumann Institute for Computing (NIC) in J\"ulich,
  at the High Performance Computing Center Stuttgart (HLRS) of the 
  University of Stuttgart, and at the Computer Center in Garching
  (RZG) are acknowledged. 
\end{theacknowledgments}




\bibliographystyle{aipproc}   

\bibliography{sample}

\IfFileExists{\jobname.bbl}{}
 {\typeout{}
  \typeout{******************************************}
  \typeout{** Please run "bibtex \jobname" to optain}
  \typeout{** the bibliography and then re-run LaTeX}
  \typeout{** twice to fix the references!}
  \typeout{******************************************}
  \typeout{}
 }

\end{document}